\documentclass[12pt]{article}
\usepackage{jheppub}
\usepackage{journals}
\usepackage[english]{babel}
\usepackage{placeins}
\usepackage{graphicx}
\usepackage{wrapfig}
\usepackage{subfig}
\usepackage{slashed}
\usepackage{amsmath}
\usepackage{amssymb}
\usepackage{feynmp-auto}
\begin {document}

\title{On high-order perturbative calculations at finite density}

\preprint{CERN-TH-2016-207, HIP-2016-25/TH}

\author[a]{Ioan Ghisoiu,}
\author[a,b]{Tyler Gorda,}
\author[c]{Aleksi Kurkela,}
\author[b,d]{Paul Romatschke,}
\author[a]{Matias S\"appi,}
\author[a]{and Aleksi Vuorinen}

\affiliation[a]{Helsinki Institute of Physics and Department of Physics, University of Helsinki, Finland}
\affiliation[b]{Department of Physics, University of Colorado Boulder, Boulder, CO, USA}
\affiliation[c]{Theoretical Physics Department, CERN, Geneva, Switzerland, and \\ Faculty of Science and Technology, University of Stavanger, Stavanger, Norway}
\affiliation[d]{Center for Theory of Quantum Matter, University of Colorado, Boulder, CO, USA}
\emailAdd{ioan.ghisoiu@helsinki.fi}
\emailAdd{tyler.gorda@helsinki.fi}
\emailAdd{aleksi.kurkela@cern.ch}
\emailAdd{paul.romatschke@colorado.edu}
\emailAdd{matias.sappi@helsinki.fi}
\emailAdd{aleksi.vuorinen@helsinki.fi}

\abstract{We discuss the prospects of performing high-order perturbative calculations in systems characterized by a vanishing temperature but finite density. In particular, we show that the determination of generic Feynman integrals containing fermionic chemical potentials can be reduced to the evaluation of three-dimensional phase space integrals over vacuum on-shell amplitudes --- a result reminiscent of a previously proposed ``naive real-time formalism'' for vacuum diagrams. Applications of these rules are discussed in the context of the thermodynamics of cold and dense QCD, where it is argued that they facilitate an extension of the Equation of State of cold quark matter to higher perturbative orders.}

\maketitle

\section{Introduction \label{sec:intro}}

Understanding the properties of cold and dense strongly interacting matter is known to be a very challenging task. With the Sign Problem preventing a lattice approach \cite{deForcrand:2010ys}, the first-principles methods available for describing, e.g., the internal composition of neutron stars, are limited to Chiral Effective Theory at low baryon densities \cite{Machleidt:2011zz} and perturbative QCD at ultrahigh density \cite{Kraemmer:2003gd}. In order to decrease the currently sizable error bars of the Equation of State (EoS) of neutron star matter \cite{Hebeler:2013nza}, it is thus clear that these two approaches should be actively pushed to higher orders. Indeed, it has been shown recently that a systematic interpolation between the low- and high-density limits can be efficiently used to restrict the behavior of the neutron star EoS at all densities, provided that the asymptotic limits are accurate enough \cite{Kurkela:2014vha,Fraga:2015xha}.

The current state-of-the-art result for the perturbative EoS of zero-temperature quark matter is from a three-loop, or ${\mathcal O}(\alpha_s^2)$, calculation that was first performed at vanishing quark masses \cite{Freedman:1976ub,Vuorinen:2003fs}, but later generalized to nonzero quark masses \cite{Kurkela:2009gj} (see also \cite{Fraga:2004gz}) and small but nonvanishing temperatures \cite{Kurkela:2016was}. Extending these zero-temperature results to higher orders, however, presents a considerable technical challenge. Similarly to the case of high temperatures \cite{Kajantie:2002wa,DiRenzo:2006nh}, part of the problem in extending these results lies in understanding how to handle the contributions of the soft momentum scales to the quantity. These difficulties will be addressed in a forthcoming publication, containing the logarithmic $\alpha_s^3\ln\,\alpha_s$ and $\alpha_s^3\ln^2\,\alpha_s$ contributions to the perturbative EoS \cite{nextpaper}.  A more challenging part of the full ${\mathcal O}(\alpha_s^3)$ result is, however, the contribution of the hard energy scale $\mu_B$, i.e.~the baryon chemical potential, which is obtained from the sum of all four-loop bubble diagrams of QCD. The high-temperature counterpart of this computation has turned out to be extremely challenging, and has only been worked out in $\phi^4$ theory \cite{Gynther:2007bw} as well as for the large-$N_f$ limit of QCD \cite{Gynther:2009qf}. 

In the paper at hand, we present a new technical tool for perturbative calculations at zero temperature but finite chemical potentials that we argue enables a high-order determination of many important thermodynamic quantities. This tool is referred to as a set of ``cutting rules'', which were proposed but not explicitly derived in ref.~\cite{Kurkela:2009gj}. They concern Feynman integrals at zero temperature and finite fermionic chemical potentials, and reduce the evaluation of the original One-Particle-Irreducible (1PI) Feynman graph to the computation of three-dimensional phase space integrals over on-shell vacuum ($T=\mu=0$) amplitudes. This represents a remarkable simplification for practical calculations, as there is a vast amount of literature on vacuum amplitudes that can be directly taken over. This significantly streamlines the evaluation of multi-loop Feynman graphs.

Although our derivation of the cutting rules utilizes the imaginary-time formalism of thermal field theory, it is interesting to note that the result appears to have an intimate connection to the real-time formalism as well. In particular, it can be shown that (modulo some simple additional assumptions) our cutting rules would emerge from a naive replacement of Euclidean propagators by the time-ordered propagators of the real-time formalism, closely reminiscent of eq.~(4) of ref.~\cite{Andersen:2000zn}. This is sometimes referred to as the ``naive real-time formalism''. This result dates back to the much earlier work of Dashen et al.~\cite{Dashen:1969ep}, where a connection between certain statistical-physics quantities and scattering amplitudes was proposed, and it has since then been developed, e.g.,~in \cite{Frenkel:1992az,Bugrii:1995vn}. It is, however, important to note that the formalism has been proposed only for vacuum diagrams, and even there no proof to all orders exists; rather, the validity of the replacement has only been checked on a case-by-case basis up to partial three-loop order. In contrast, our proof of the zero-temperature cutting rules covers all Euclidean $n$-point functions up to an arbitrary order in perturbation theory, thereby validating the use of the naive real-time formalism for these quantities.

Our paper is organized as follows. In section \ref{sec:cut}, we introduce our notation and state the cutting rules. In addition, as an illustration, we present a simple two-loop computation in two ways: both without and with the help of the cutting rules. Section \ref{sec:proof} then contains a detailed proof of the rules, as well as two intermediate lemmas that are each interesting in their own right. After this, we discuss the regularization of the most common divergences occurring in the cut graphs in section \ref{sec:reg}, while section \ref{sec:conc} presents an outlook on the applications of the cutting rules, in particular in dense QCD. Lastly, many details of the more subtle parts of our proof have been relegated to the appendixes.

\section{Cutting rules \label{sec:cut}}

\subsection{Notation and statement of the rules \label{sec:state}}

We work with Euclidean signature Feynman graphs at zero temperature and finite chemical potentials. This means that we can think of our diagrams as consisting of two types of propagators, ``fermionic'' $1/((q_0+i\mu)^2+(E_q^i)^2)$ and ``bosonic'' $1/(q_0^2+(E_q^i)^2)$, where $E_q^i\equiv \sqrt{q^2+m_i^2}$ and $\mathbf{q}$ represents a spatial momentum vector. Consistently with the fermionic nature of the chemical potential, we assume $\mu$ to be larger than the mass of the corresponding field.  
Divergences are finally regulated via dimensional regularization by working in $d=3-2\epsilon$ spatial dimensions, defining
\begin{eqnarray}
\int_Q\,\equiv\, \int_{-\infty}^\infty \frac{dq_0}{2\pi}\int_q \,\equiv\,\int_{-\infty}^\infty \frac{dq_0}{2\pi} \int\frac{d^{d}q}{(2\pi)^{d}}\, ,
\end{eqnarray}
where $Q$ denotes a Euclidean four-vector, such that $Q^2\equiv q_0^2+q^2$. 

Before stating the cutting rules, we make a few simplifying assumptions, the purpose of which is to keep our presentation as concise and readable as possible:
\begin{itemize}
\item There is no structure in the numerator of the original Feynman integral, i.e.~we consider scalar propagators and trivial vertex functions.
\item No individual propagator is raised to a power higher than one.
\item There is only one chemical potential appearing in the graph.
\item In the external momenta $P_k=(p_0^k,\mathbf{p}_k)$, the $p_0^k$ are always real-valued, corresponding to imaginary frequencies $\omega_k$.
\end{itemize}
As will be discussed in section \ref{sec:conc}, the first three of these assumptions can be easily relaxed, but at the cost of making the notation somewhat more convoluted. Note, however, that we have made absolutely no assumptions about the masses of the propagators, so that they can and will be considered independent.

Under the above assumptions, let us consider an arbitrary 1PI $N$-loop $n$-point Feynman graph  $F(\{P_k\},\mu)$, where the $P_k$, $k=1,2,...,n$ stand for the external momenta. According to the cutting rules, we may write this function in the form
\begin{eqnarray}
F(\{P_k\},\mu)&=&F_\text{0-cut}(\{P_k\})+F_\text{1-cut}(\{P_k\},\mu)+\cdots+F_\text{$N$-cut}(\{P_k\},\mu), 
\end{eqnarray}
where  $F_\text{0-cut}(\{P_i\})$ is simply the original graph evaluated at vanishing chemical potential, $\mu=0$, while the remaining pieces result from the cutting procedure. In particular, $F_\text{$j$-cut}(\{P_k\})$ denotes the sum of all so-called ``$j$-cut'' diagrams, in which exactly $j$ of the internal fermionic propagators have been cut off. This cutting procedure involves the following steps: 
\begin{enumerate}
 \item Removing the cut propagators from the original graph.
 \item Evaluating the resulting $N-j$ -loop $n+2j$ -point amplitude at $T=\mu=0$, assuming all external momenta to be real-valued.
 \item Setting the cut momenta $Q_i$ on shell, i.e.~writing $q_0^i=iE_i$ for each of them.
 \item Integrating the resulting expression over the cut three-dimensional momenta with the weights $-\theta(\mu-E_i)/(2E_i)$.
\end{enumerate}
An important additional rule is that those cuts that divide the original graph into two or more disconnected pieces are to be thrown out.

The usefulness of the cutting rules originates from the fact that they isolate the chemical-potential dependence of the original graph in the $\theta$-function factors in the three-dimensional ``phase space'' integrations. Owing to the abundance of analytic results for vacuum amplitudes in the literature, one typically only needs to perform (some of) these phase space integrations numerically, which is an enormous simplification.

\subsection{Example calculation: standard technique}

\begin{figure}[t]
\begin{center}
\includegraphics[width=.8\linewidth]{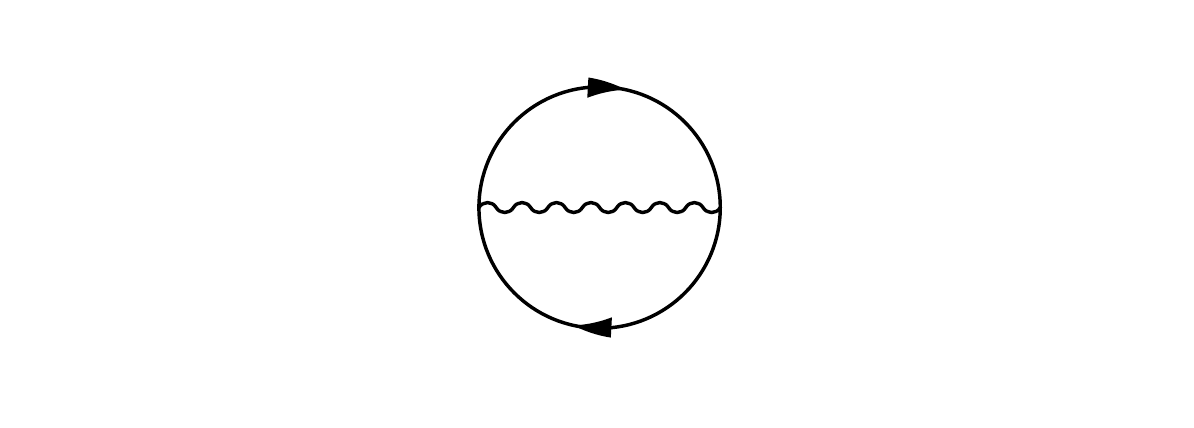}
\caption{\label{fig:mot1} A two-loop diagram contributing to the partition functions of QED and QCD, with the wiggly line corresponding to a gauge boson (photon or gluon) and the solid lines to fermions (electrons or quarks).}
\end{center}
\end{figure}

To illustrate the use of the cutting rules as well as motivate their introduction, we consider next an elementary practical example in the form of a two-loop integral encountered when considering the graph of fig.~\ref{fig:mot1}, appearing in the ${\mathcal O}(\alpha_s)$ contribution to the EoS of QED or QCD matter. At zero temperature, the most nontrivial master integral resulting from this diagram is the two-loop ``sunset'' 
\begin{fmffile}{2loop}
\begin{eqnarray}
\label{eq:abc2}
\!\!I_2(\mu)&\equiv&\!\!\!\!\!\!
\begin{gathered}
			\begin{fmfgraph*}(65,60)
				\fmfleft{in1}
				\fmfright{out1}
				\fmf{phantom,tension=5}{in1,vin1}
				\fmf{phantom,tension=5}{vout1,out1}
				\fmf{plain,left,tension=0.4}{vin1,vout1,vin1}
				\fmf{dots,width=2}{vin1,vout1}
			\end{fmfgraph*}
		\end{gathered}
 \!\!\!\!\!=\int_P \int_Q \frac{1}{(p_0+i\mu)^2+E_p^2}\frac{1}{(q_0+i\mu)^2+E_q^2}\frac{1}{(p_0-q_0)^2+(\mathbf{p}-\mathbf{q})^2} \\
&=&\int_P \int_Q \int_K \frac{(2\pi)^4\delta^{(4)}(P-Q-K)}{\left((p_0+i\mu)^2+E_p^2\right)\left((q_0+i\mu)^2+E_k^2\right)\left(k_0^2+k^2\right)}\,,\quad E_k\equiv\sqrt{k^2+m^2}\,, \nonumber
\end{eqnarray}
where the solid lines in the graph denote a massive fermion propagator and the dotted line a massless boson one. In order to reduce the integral to a more manageable form, we first perform the integrations over the 0-components of the different momenta, which is a rather straightforward task at such a low loop order.

The standard way of evaluating the $p_0$- and $q_0$-integrals proceeds by writing the $\delta$-function in eq.~(\ref{eq:abc2}) in the form \cite{Kapusta:2006pm}
\begin{equation}
\label{eq:delta}
2\pi \delta(p_0-q_0-k_0)=\int_{-\infty}^\infty d\alpha\, e^{i\alpha(p_0-q_0-k_0)}\,,
\end{equation}
which allows us to  perform the $p_0$, $q_0$, and $k_0$ integrations independently using the Residue theorem. Choosing the integration contours to lie on the upper or lower half of the complex plane depending on the sign of the exponent in eq.~(\ref{eq:delta}), we obtain after quite some algebra
\begin{eqnarray}
I_2(\mu)&=&\int_{p,q}\frac{1}{2|\mathbf{p}-\mathbf{q}|\,2E_p \,2E_q}\frac{2}{|\mathbf{p}-\mathbf{q}|+E_p+E_q}\nonumber\\
&&-\int_{p,q}\frac{\theta(\mu-E_p)}{2|\mathbf{p}-\mathbf{q}|\, 2E_p\, 2E_q}\frac{2 (|\mathbf{p}-\mathbf{q}|+E_q)}{(|\mathbf{p}-\mathbf{q}|+E_q)^2-E_p^2}\nonumber\\
&&-\int_{p,q}\frac{\theta(\mu-E_q)}{2|\mathbf{p}-\mathbf{q}|\, 2E_p\, 2E_q}\frac{2 (|\mathbf{p}-\mathbf{q}|+E_p)}{(|\mathbf{p}-\mathbf{q}|+E_p)^2-E_q^2}\nonumber\\
&&+\int_{p,q}\frac{\theta(\mu-E_p) \theta(\mu-E_q)}{2E_p\,2E_q}  \frac{1}{(\mathbf{p}-\mathbf{q})^2-(E_p-E_q)^2}\, . \label{I2int} 
\end{eqnarray}
While perfectly correct, this result is unfortunately rather unpractical, as the first three lines contain complicated UV divergent integrals of highly non-standard objects that we would need to determine in $3-2\epsilon$ dimensions. Only the last of the four terms is of a form that may be directly evaluated as a numerical integral. 

Some insights into how the above integrals might become tractable can be gained by observing that the first term of eq.~(\ref{I2int}), containing no step functions, may be recast in the form of the original diagram evaluated at $\mu=0$,
\begin{eqnarray}
&&\int_{p,q}\frac{1}{2|\mathbf{p}-\mathbf{q}|\,2E_p \,2E_q}\frac{2}{|\mathbf{p}-\mathbf{q}|+E_p+E_q} \nonumber \\
&=& \int_P \int_Q \frac{1}{\left(p_0^2+E_p^2\right)\left(q_0^2+E_k^2\right)\left((p_0-q_0)^2+(\mathbf{p}-\mathbf{q})^2\right)} = I_2(0).
\label{eq:mot1}
\end{eqnarray}
Apart from being of limited physical interest, this term can be easily evaluated using the standard tools of perturbative zero-temperature field theory. 

Perhaps even more interestingly, we note that the second term of  eq.~(\ref{I2int}) is expressible in terms of a two-point function,
\begin{eqnarray}
 \int_Q \frac{1}{q_0^2+E_q^2}\frac{1}{(p_0-q_0)^2+(\mathbf{p}-\mathbf{q})^2}\Big{|}_{p_0\to iE_p}=\int_q\frac{1}{2|\mathbf{p}-\mathbf{q}|\,2E_q}\frac{2 (|\mathbf{p}-\mathbf{q}|+E_q)}{(|\mathbf{p}-\mathbf{q}|+E_q)^2-E_p^2},\nonumber \\ \label{eq:I21}
\end{eqnarray}
where the substitution $p_0\to iE_p$ is made only after performing the integral on the left-hand side (prior to this substitution, $p_0$ is considered real). Again, we note that the integral on the left-hand side is of a form often encountered in standard $T=0$ quantum field theory calculations. A similar result clearly exists for the third term of  eq.~(\ref{I2int}) as well, via the substitution $\mathbf{p}\leftrightarrow\mathbf{q}$.

\subsection{Example calculation: cutting rules}

The observations made in the previous subsection clearly suggest that the two-loop integral of eq.~(\ref{eq:abc2}) 
may be written in a form reminiscent of the cutting rules, i.e.~as a sum of phase-space integrals of higher-point amplitudes. To make this statement more precise, we shall now explicitly show that the cutting rules indeed exactly reproduce the above results. 

According to the cutting rules, the integral $I_2(\mu)$ of eq.~(\ref{eq:abc2}) can be directly written in the  form
\begin{eqnarray}
 I_2(\mu)&\equiv& I_2^\text{0-cut}+ I_2^\text{1-cut}(\mu)+I_2^\text{2-cut}(\mu)\,.
  \end{eqnarray}
Here, the first term reads
\begin{eqnarray}
I_2^\text{0-cut}&=& I_2(0)\:=\!\!\!\!
		\begin{gathered}
			\begin{fmfgraph*}(60,60)
				\fmfleft{in2}
				\fmfright{out2}
				\fmf{phantom,tension=5}{in2,vin2}
				\fmf{phantom,tension=5}{vout2,out2}
				\fmf{dashes,left,tension=0.4}{vin2,vout2,vin2}
				\fmf{dots,width=2}{vin2,vout2}
			\end{fmfgraph*}
		\end{gathered}\!\!\!\!,
\end{eqnarray}
where we have denoted a $\mu=0$ massive propagator by a dashed line. This clearly agrees with eq.~(\ref{eq:mot1}).
For the one-cut piece, on the other hand, we obtain
\begin{eqnarray}
I_2^\text{1-cut}(\mu)&=& -\int_p \frac{\theta(\mu-E_p)}{2E_p} \Bigg[\int_Q \frac{1}{q_0^2+E_q^2}\frac{1}{(p_0-q_0)^2+(\mathbf{p}-\mathbf{q})^2}\Bigg]_{p_0\to iE_p} \nonumber \\
&-&\int_q \frac{\theta(\mu-E_q)}{2E_q} \Bigg[\int_P \frac{1}{p_0^2+E_p^2}\frac{1}{(p_0-q_0)^2+(\mathbf{p}-\mathbf{q})^2}\Bigg]_{q_0\to iE_q} \nonumber 
\\
&\equiv&
	-\:2\intop_{p}\frac{\theta(\mu - E_{p})}{2 E_{p}}\;
	\begin{gathered}
			\begin{fmfgraph*}(60,60)
				\fmfleft{in3}
				\fmfright{out3}
				\fmf{dashes,tension=5}{in3,vin3}
				\fmf{dashes,tension=5}{vout3,out3}
				\fmf{dashes,left,tension=0.4}{vin3,vout3}
				\fmf{dots,width=2}{vin3,vout3}
			\end{fmfgraph*}
		\end{gathered}\,
		\Bigg{|}_{p_{0}\rightarrow iE_{p}}\,, \label{sunset1}
\end{eqnarray}
where we have used the symmetry of the two terms in the first form of the result. This can be identified with the second and third terms of eq.~(\ref{I2int}), using eq.~(\ref{eq:I21}). Finally, the two-cut part of the graph takes the form
\begin{eqnarray}
I_2^\text{2-cut}(\mu)&=& \int_p \frac{\theta(\mu-E_p)}{2E_p}\int_q \frac{\theta(\mu-E_q)}{2E_q} \Bigg[\frac{1}{(p_0-q_0)^2+(\mathbf{p}-\mathbf{q})^2}\Bigg]_{p_0\to iE_p,\,q_0\to iE_q}\;\;\nonumber
\\
&\equiv&
	\intop_{p}\frac{\theta(\mu - E_{p}) }{2 E_{p}}
			\intop_{q}\frac{\theta(\mu - E_{q}) }{2 E_{q}}
		\begin{gathered}
			\begin{fmfgraph*}(80,40)
				\fmfleft{ind4,inu4}
				\fmfright{outd4,outu4}
				\fmf{dashes,tension=0.3}{ind4,vin4,inu4}
				\fmf{dashes,tension=0.3}{outd4,vout4,outu4}
				\fmf{dots,width=2,tension=0.2}{vin4,vout4}
			\end{fmfgraph*}
		\end{gathered}		
			\Bigg{|}_{p_{0}\rightarrow iE_{p},q_{0}\rightarrow iE_{q}}, \label{sunset2}
\end{eqnarray}
which is nothing but the last term of eq.~(\ref{I2int}). Thus, the cutting rules do indeed reproduce the full result for the graph considered. It is worth pointing out here that this result is by no means new and only serves as a pedagogical introduction to our discussion; in a finite-$T$ context, a similar result has been obtained, e.g.,~in Appendix A of \cite{Laine:2006cp}.
\end{fmffile}

\section{Proof of the rules \label{sec:proof}}

\subsection{Organization of the proof}

In this section, we provide a proof of the cutting rules in the case of a generic 1PI Feynman graph at zero temperature but finite chemical potential. For reasons of clarity, we do this in three sequential steps, of which the first two can be considered useful lemmas while the connection to the cutting rules is made in the third part of the proof:
\begin{enumerate}
 \item \textit{Vanishing-chemical-potential case}: Consider a generic Euclidean Feynman integral at zero temperature and vanishing chemical potential, corresponding to a 1PI $N$-loop $n$-point diagram. Such a graph consists of some (potentially large) number $M$ of internal lines, which we enumerate by the index $\alpha=1,2,...,M$. The corresponding propagators can be written in the form
 \begin{eqnarray}
  \frac{1}{(r_0^\alpha)^2+E_\alpha^2}, \quad E_\alpha\equiv\sqrt{r_\alpha^2+m_\alpha^2}\, ,
 \end{eqnarray}
of which exactly $N$ can be chosen to correspond to the loop momenta $Q_i$, $i=1,...,N$, that are integrated over. The remaining $M-N$ $R_\alpha$ are then linear combinations of the loop momenta and the external momenta $P_k$, $k=1,...,n$, as dictated by momentum conservation at the vertices (see appendix \ref{sec:mom} for a discussion of this issue).

Our claim is now that the integral
  \begin{eqnarray}
 I&\equiv& \int_{-\infty}^\infty \prod_{i=1}^{N} \frac{dq_0^i}{2\pi}\, \prod_{\alpha=1}^M \frac{1}{(r_0^\alpha)^2+E_{\alpha}^2}, \label{Ires}
 \end{eqnarray}
where we have made some arbitrary choice for the loop momenta, can be written in a simple form that is explicitly independent of this choice. To write down this result, we introduce the following notation: denoting the set of all propagators by $P\equiv\{1,2,...,M\}$, we define $S$ to be the set of all possible choices of loop momenta, such that each element $S_r\in S$ corresponds to some subset of $N$ indices from $P$. The sets $S_r$ are limited only by momentum conservation, and several examples of them are given in appendix \ref{sec:mom}. 

With the above notation, our proposed result for $I$ reads
\begin{eqnarray}
I&=&\sum_{S_r\in S}\;\prod_{i\in S_r} \frac{1}{2E_i} \prod_{\alpha\in P\setminus S_r} \frac{1}{(r_0^\alpha(S_r))^2+E_\alpha^2(S_r)}\Big{|}_{\{q_0^i=iE_i\}}, \label{Iresmu0}
\end{eqnarray}
where $P\setminus S_r$ denotes the propagators that do not belong to the set $S_r$ and the explicit forms of the $R_\alpha$ in terms of the $Q_i$ and $P_k$ are dictated by $S_r$. Note that each set $S_r$ is to be counted only once here, and the labeling of the momenta within $S_r$ (including changing the direction of some loop momenta, $Q_i\to-Q_i$) plays no role: it is only the choice of propagators that counts.

\item \textit{Generalization to finite density}: Assume next that some of the internal propagators of the graph are fermionic in the sense that they carry a chemical potential $\mu$ in the way stated in the previous section. The above result for $I$ is then modified only by the factors $1/(2E_i)$ corresponding to the internal fermion lines changing according to
\begin{eqnarray}
\frac{1}{2E_i}&\to& \frac{\theta(E_i-\mu)}{2E_i} \, .
\end{eqnarray}
Note that for different $S_r\in S$, the numbers of fermionic momenta and thus $\theta$-function factors are typically different.

\item \textit{Connection to the original cutting rules}: Writing the $\theta$-functions in the form $\theta(E_i-\mu) = 1-\theta(\mu-E_i)$ and rearranging terms, the above results reduce to the cutting rules stated in section \ref{sec:cut}.
\end{enumerate}
We now proceed to prove these three claims, thereby deriving the cutting rules.

\subsection{Vanishing-chemical-potential case}

Given a random choice of loop momenta $S_r\in S$, we may clearly express the integral we wish to evaluate [cf.~eq.~(\ref{Ires})] in the form
\begin{eqnarray}
I(S_r)&\equiv& \int_{-\infty}^\infty \prod_{i \in S_r}\frac{dq_0^i}{2\pi}\, \prod_{\alpha=1}^M \frac{1}{(r_0^\alpha(S_r))^2+E_{\alpha}^2(S_r)}\,, \label{Iresx}
\end{eqnarray}
where our notation highlights the fact that the $R_\alpha$ and corresponding $E_\alpha$ depend on $S_r$. Picking some $i \in S_r$ as the first integration to be performed gives a sum of residues of the form $\frac{1}{2E_{\alpha}}\left(\ldots\right)|_{q_0^i = iE_\alpha+...}$, where each $\mathbf{r}_\alpha$ depends linearly on the $\mathbf{q}_i$. Placing the momentum in question on shell shifts the poles of some of the remaining propagators but does not affect the corresponding residues. Upon subsequent integrations, complicated combinations of $\theta$-functions typically appear in the numerator of the result due to these shifts. What remains unchanged, however, is that each integration produces an additional factor of $1/(2E_{\alpha'})$, originating from the residue of one of the remaining propagators. 

To illustrate the above reasoning, consider the following simple example, where we perform three $q_0$ integrations, always picking up the pole from the highlighted propagator:
\begin{eqnarray}
&&\int_{q_0^1}\int_{q_0^2}\int_{q_0^3}\mathbf{\frac{1}{(q_0^1)^2+E_1^2}}\frac{1}{(q_0^2)^2+E_2^2}\frac{1}{(q_0^3)^2+E_3^2} \frac{1}{(q_0^1+q_0^2)^2+E_4^2}\frac{1}{(q_0^1+q_0^2-q_0^3)^2+E_5^2} \nonumber \\
&=& \int_{q_0^2}\int_{q_0^3}\frac{1}{2E_1}\frac{1}{(q_0^2)^2+E_2^2}\frac{1}{(q_0^3)^2+E_3^2}\mathbf{\frac{1}{(iE_1+q_0^2)^2+E_4^2}}\frac{1}{(iE_1+q_0^2-q_0^3)^2+E_5^2}+\cdots \nonumber \\
&=& \int_{q_0^3}\frac{1}{2E_1}\frac{\theta(E_4-E_1)}{2E_4}\frac{1}{(iE_4-iE_1)^2+E_2^2}\frac{1}{(q_0^3)^2+E_3^2}\mathbf{\frac{1}{(iE_4-q_0^3)^2+E_5^2}} +\cdots\nonumber \\
&=& \frac{1}{2E_1}\frac{\theta(E_4-E_1)}{2E_4}\frac{\theta(E_5-E_4)}{2E_5}\frac{1}{(iE_4-iE_1)^2+E_2^2}\frac{1}{(iE_4-iE_5)^2+E_3^2} +\cdots\, .
\end{eqnarray}
Note that additional terms of the exact same form but with different $\theta$-function factors originate from taking the same poles in a different order.

From the above exercise, we see that $I(S_r)$ obtains the form of a lengthy sum of terms of a similar kind: a product of residues $1/(2E_\alpha)$ from some set ${\mathcal S}_s$ of $N$ propagators, multiplied by a complicated sum of products of $\theta$ functions along with the product of the remaining propagators, with the ${\mathcal S}_s$ momenta placed on shell. Defining a function $A_{S_r}({\mathcal S}_s)$ to stand for the latter part of the result, we may write it in the form
\begin{eqnarray}
I(S_r)&=&\sum_{{\mathcal S}_s} \prod_{k\in {\mathcal S}_s}\frac{1}{2E_k}\,A_{S_r}({\mathcal S}_s)\Big{|}_{\{q_0^k=iE_k\}}.
\end{eqnarray}
A crucial realization is now the following: the sets of $N$ propagators obtained above, i.e.~the ${\mathcal S}_s$, cannot contain any sets that are not part of the ``superset'' $S$, defined as all the possible choices of loop momenta in the original graph. This is a simple consequence of momentum conservation: we cannot take residues of a set of propagators whose momenta are linearly dependent. This implies that we may directly write the result in a form reminiscent of eq.~(\ref{Iresmu0}),
\begin{eqnarray}
I(S_r)&=&\sum_{S_{r'}\in S}\;\prod_{i\in S_{r'}} \frac{1}{2E_i}\, \tilde{A}_{S_r}(S_{r'})\Big{|}_{\{q_0^i=iE_i\}}, \label{Iresmu1}
\end{eqnarray}
where the tilde on $A$ highlights the fact that the summation now goes over the sets $S_{r'}$.

The remaining step in relating the above result to eq.~(\ref{Iresmu0}) is to use the known independence of $I(S_r)$ on the random initial set $S_r$, i.e.~the fact that $I(S_{r})=I$. For any given $S_r\in S$, there is one term in the sum of eq.~(\ref{Iresmu1}) that is particularly simple, namely that where $S_{r'}=S_{r}$. For this term, each of the $q_0^i$ integrations picks up a pole from a propagator of the simple form $1/((q_0^i)^2+E_i^2)$, so that no $\theta$-functions arise, producing
\begin{eqnarray}
\tilde{A}_{S_r}(S_{r})&=& \prod_{\alpha\in P\setminus S_r} \frac{1}{(r_0^\alpha(S_r))^2+E_\alpha^2(S_r)}\, . \label{ASr}
\end{eqnarray} 
Owing to the independence of the $E_\alpha$, we on the other hand know that the different terms in the sum of eq.~(\ref{Iresmu1}) must be unique (see appendix \ref{sec:ind} for a detailed discussion of this point), so that $\tilde{A}_{S_r}(S_{r'})=\tilde{A}_{S_{r'}}(S_{r'})$ for all $S_r, S_{r'}$. This implies that also the coefficients $\tilde{A}_{S_r}(S_{r'})$, $r\neq r'$, must reduce to the simple form
\begin{eqnarray}
\tilde{A}_{S_r}(S_{r'})&=& \prod_{\alpha\in P\setminus S_{r'}} \frac{1}{(r_0^\alpha(S_{r'}))^2+E_\alpha^2(S_{r'})}\,, \label{ASr2}
\end{eqnarray}
which --- together with eq.~(\ref{Iresmu1}) --- leads us directly to the result (\ref{Iresmu0}).

\subsection{Generalization to finite density}

Somewhat surprisingly, the generalization of the above result to the presence of nonzero $\mu$ in some of the propagators is by far the simplest part of our proof. Namely, the exact same reasoning goes through as in the $\mu=0$ case, with the only modification being a shift in the poles and residues of the fermion propagators originating from the $\mu$-dependence of the integral
\begin{eqnarray}
\int_{-\infty}^\infty dq_0^i\, \frac{1}{(q_0^i+i\mu)^2+E_i^2}&=&\frac{\theta(E_i-\mu)}{2E_i}\, .
\end{eqnarray}
In other words: whenever the pole of a fermionic propagator is taken, we need to multiply the corresponding residue in the product $\prod_{i\in S_{r'}} \frac{1}{2E_i}$ by the factor $\theta(E_i-\mu)$.

\subsection{Connection to the original cutting rules}

The previous step of the proof brought us to a somewhat cumbersome result, featuring a sum over all possible labelings of loop momenta in the original Feynman graph, with each term in the sum further containing a product of some number of $\theta(E_i-\mu)$ factors. To move forward, we write each of the $\theta$-functions in the form $1-\theta(\mu-E_i)$, and then reassemble the result for $I$ in the form
\begin{eqnarray}
 I&=& \Big(\text{terms with 0 $\theta(\mu-E_i)$'s}\Big)+\Big(\text{terms with 1 $\theta(\mu-E_i)$}\Big)+\cdots  \nonumber \\
 &&+  \Big(\text{terms with $N$ $\theta(\mu-E_i)$'s}\Big)\, .
\end{eqnarray}
It is self-evident that the first term in the above sum corresponds to the $\mu=0$ version of the same graph, but a little more effort is required to see what happens to the terms with one or more $\theta$-functions. 

In the single-$\theta$ part of the result, we first group together terms according to the argument of the $\theta(\mu-E_i)$ function they contain, which clearly correspond to all the fermionic propagators in the original graph. Singling out one of them (and the associated $-\frac{1}{2E_i}$ factor), we note that it is multiplied by a sum of terms, each of which contains a product of $N-1$ factors of $1/(2E_j)$ as well as the product of the rest of the propagators with the $E_i$ and $E_j$ lines placed on shell. Recalling the result of the first part of our proof, we recognize this as the result for an $N-1$ -loop $n+2$ -point function that is obtained by removing the line $i$ from the original graph, i.e.
\begin{eqnarray}
I&=&\cdots -\frac{\theta(\mu-E_i)}{2E_i} \times \Big(\text{original graph with $Q_i$-propagator removed}\Big)\Big{|}_{q_0^i=iE_i} + \cdots \, , \nonumber \\
\end{eqnarray}
so that the sum of all such terms exactly corresponds to the sum of all 1-cut graphs in the cutting rules. In evaluating this expression, the $Q_i$ line clearly must be placed on-shell only after computing the associated $n+2$ -point function, as one of the assumptions of the $\mu=0$ cutting rules above was that all external momenta in the original graph be real-valued (modulo a possible $\mu$ in the external legs of the original graph).

Moving on to the terms with two or more $\theta$-function factors, the above reasoning goes through in each case, leaving us with the 2-, 3-, ..., and $N$-cut contributions to the original graph. In each case, the generated amplitudes are to be evaluated assuming the external momenta to be real: only afterwards are the cut momenta placed on-shell.

One final comment is in order. Each time some number of fermion lines are cut in a given Feynman graph, it follows from the construction presented above that these propagators must form a subset of some possible choice of integration momenta $S_r\in S$. This means that the cuts can never split the original 1PI graph into two (or more) disconnected pieces: for this to happen, we would need to cut propagators whose momenta are not linearly independent, which is not possible for any subset of $S_r$. 

\section{Regularization of the integrals \label{sec:reg}}

Before we can successfully apply the cutting procedure to an arbitrary Feynman diagram, there is one further issue that needs to be discussed. This is related to the regularization of unphysical divergences appearing in the calculations, of which there are two distinct variations. They differ in that the first kind of divergence appears in the very definition of the finite-$\mu$ integrals, while the latter is a byproduct of the cutting procedure and therefore more artificial.

The first type of singularity is related to the divergence of the fermionic propagator $1/((p_0+i\mu)^2+E_p^2)$ when $p_0=0$ and $E_p=\mu$, i.e.~it appears along the original integration contour. It gets realized only in the limit where the $\theta(\mu-E_p)$ function in the integration measure gets saturated, but one might nevertheless worry that it makes the $p_0$ integrations ill-defined. The most natural resolution to this problem turns out to involve the use of an infinitesimal but nonzero temperature $T$ as a regulator. As we shall show in detail in appendix \ref{sec:matsu}, the effect of this prescription amounts to interpreting all $p_0$ integrations in the principal value sense, i.e.~writing
\begin{eqnarray}
\int_{-\infty}^\infty \frac{dp_0}{2\pi}\int_p &\to&\mathcal{P}\int_{-\infty}^\infty \frac{dp_0}{2\pi}\int_p\;=\; \lim_{\epsilon\to 0^+}\Bigg\{\int_{-\infty}^{-\epsilon} \frac{dp_0}{2\pi}+\int_{\epsilon}^{\infty} \frac{dp_0}{2\pi}\Bigg\}\int_p \nonumber \\
&=&\frac{1}{2}\,\Bigg\{\int_{-\infty+i0^+}^{\infty+i0^+} \frac{dp_0}{2\pi}+\int_{-\infty-i0^+}^{\infty-i0^+} \frac{dp_0}{2\pi}\Bigg\}\int_p\,, \label{principal}
\end{eqnarray}
where we have assumed $p_0=0$ to be the only singular point on the real axis. While this does not affect the practical application of the cutting rules, it demonstrates that the integrand is well-defined on the entire integration contour, so that no imaginary parts can be generated in bubble graphs due to the divergence.

Another frequently occurring problem is the emergence of spurious poles in some of the cut graphs that would automatically cancel, should all of the $p_0$ integrations in the diagram be computed at the same time and the results added together. A simple example of this is seen in the integral
\begin{eqnarray}
&&\int_{-\infty}^\infty\frac{dp_0}{2\pi}\int_{-\infty}^\infty\frac{dq_0}{2\pi}\frac{1}{p_0^2+E_1^2}\frac{1}{q_0^2+E_2^2}\frac{1}{(p_0-q_0)^2+E_3^2}\nonumber \\
&=& \frac{1}{2E_1}\frac{1}{2E_2}\frac{1}{(iE_1-iE_2)^2+E_3^2}+\frac{1}{2E_1}\frac{1}{2E_3}\frac{1}{(iE_1+iE_3)^2+E_2^2}+\frac{1}{2E_2}\frac{1}{2E_3}\frac{1}{(iE_2+iE_3)^2+E_1^2}\nonumber \\
&=&\frac{1}{4E_1 E_2 E_3(E_1+E_2+E_3)}\,,
\end{eqnarray}
where the intermediate stage corresponds to the outcome of the cutting rules. Even though the initial integral as well as its final form are both perfectly well-defined for all real-valued $E_i$, we see that the intermediate result contains a sum of three terms that each diverge when the three energies satisfy the linear relation $E_1-E_2=\pm E_3$. This is clearly a deeply unphysical problem.

The simplest manifestation of the second type of divergence is seen in the two-loop sunset graph, introduced already in sec.~\ref{sec:cut}, where we now set the mass of one of the two fermion lines to zero. Considering the two-cut contribution corresponding to eq.~(\ref{sunset2}), we are left to evaluate the integral
\begin{eqnarray}
\!\!\!\!I_2^\text{2-cut}(\mu)= \int_p \frac{\theta(\mu-E_p)}{2E_p}\int_q \frac{\theta(\mu-q)}{2q} \Bigg[\frac{1}{(p_0-q_0)^2+(\mathbf{p}-\mathbf{q})^2}\Bigg]_{p_0\to iE_p,\,q_0\to iq}\,, \label{pvex}
\end{eqnarray}
where the integrand
\begin{eqnarray}
\Bigg[\frac{1}{(p_0-q_0)^2+(\mathbf{p}-\mathbf{q})^2}\Bigg]_{p_0\to iE_p,\,q_0\to iq}&=&\frac{1}{2E_p\, q-2\,\mathbf{p}\cdot\mathbf{q}-m^2}
\end{eqnarray}
contains a singularity that cannot be regulated using dimensional regularization.
For bubble diagrams and those $N$-point functions that are known to be real-valued, the choice of regulator is in principle free, but the most straightforward prescription is to interpret the diverging integrations in a principal value sense. For correlators that might develop physical imaginary parts upon a specific $i\epsilon$ prescription, the procedure is, however, more tricky and one must be careful not to discard any physically meaningful contributions.

\section{Discussion and outlook \label{sec:conc}}

The cutting rules we stated and proved in the previous three sections apply as such only under the assumptions listed in sec.~\ref{sec:state}. It is, however, straightforward to see that the first three of the conditions can be easily relaxed:
\begin{itemize}
 \item The appearance of external or internal momenta in the numerator of the Feynman graph does not prohibit the application of the cutting rules. The only potential problem might originate from 0-components of internal momenta, but even these will simply be replaced by the corresponding $iE_i$ factors according to the Residue Theorem.
 \item If a scalar propagator is raised to a higher power, care must be taken when evaluating the Feynman integral. The most straightforward way to proceed is by first evaluating the corresponding graph with the propagator raised to power 1, and then (repeatedly) differentiating the result with respect to the mass squared of the propagator in question, relying on the formula
 \begin{eqnarray}
  \frac{1}{(Q^2+m_i^2)^n}&=&\frac{(-1)^{n-1}}{(n-1)!}\frac{d}{dm_i^2}  \frac{1}{Q^2+m_i^2}\,.
 \end{eqnarray}
A possible caveat here has to do with massless propagators raised to higher powers and the associated physical IR divergences. If we introduce a mass term for such a line and then differentiate with respect to it, this will in general produce a $1/m_i^k$ term in the $m_i\to 0$ limit. Some extra effort will then be required to convert this divergence into a $1/\epsilon$ pole, as expected in dimensional regularization.
 \item Having several closed fermion loops in the graph, each with an independent chemical potential, clearly produces a mere notational complication, and the form of the result stays exactly the same as above. We only need to keep track of the correspondence of the chemical potentials with the cut fermion lines.
\end{itemize}
Together, these three generalizations allow us to tackle all Feynman integrals encountered in gauge field theories coupled to Dirac fermions, such as QED or QCD.

As discussed already in section \ref{sec:intro}, the cutting rules become increasingly important when one tries to extend perturbative studies of the thermodynamics of cold and dense systems to higher loop orders. The rules were an integral part of the determination of the three-loop EoS of cold quark matter in ref.~\cite{Kurkela:2009gj}, and it is because of them that an extension of this result to the full four-loop order is feasible. In this context, there are in fact two separate challenges: in addition to the evaluation of all four-loop bubble diagrams one needs to determine (a specific component of) the gluon polarization tensor to two-loop order. The latter of these two computations is alone sufficient for determining the logarithmic contributions $\alpha_s^3\ln^2\,\alpha_s$ and $\alpha_s^3\ln\,\alpha_s$ to the EoS. This work is near completion, and the results will be presented in a separate publication later \cite{nextpaper}. 

To conclude, let us briefly return to the connection between our work and the naive real-time formalism discussed in sec.~\ref{sec:intro}. In proving the validity of the zero-temperature cutting rules, we have, in effect, also shown that the naive real-time formalism is applicable not only for vacuum diagrams, contributing to the free energy, but also for Euclidean $n$-point functions in the $T=0$ limit. At the same time, we know that the multitude of various Minkowskian correlators (retarded, advanced, time-ordered, etc.) at nonzero temperature can only be reproduced using the Feynman rules of the full real-time formalism, featuring, in particular, a doubling of field variables. Trying to gain a detailed understanding of the conditions, under which the full real-time formalism reduces to its naive version, is clearly an intriguing avenue for future research.

\section*{Acknowledgments}

IG and AV were supported in part by the Academy of Finland, grant no.~1273545 and 1303622. PR was supported in part by the Department of Energy, DOE, award no.~DE-SC0008132.

\appendix

\section{On the choice of loop momenta \label{sec:mom}}

When expressing an $N$-loop Feynman diagram in momentum space, there are a number of possible choices for the integration (or loop) momenta. We choose each of them to correspond to the momentum flowing along one of the propagators, in which case their assignment is limited by two rules, both related to momentum conservation: 
\begin{enumerate}
 \item All internal lines meeting at a given vertex or subdiagram cannot be chosen to correspond to independent loop momenta, as they are linearly dependent.
 \item For each closed loop in the graph, at least one of the propagators forming the loop must be chosen to correspond to a loop momentum.
\end{enumerate}
Besides these rules, the choice is arbitrary, and each choice merely corresponds to a slightly different way of writing the original graph. However, they must all lead to the same result.

\begin{figure}[t]
\begin{center}
\includegraphics[width=10.5cm]{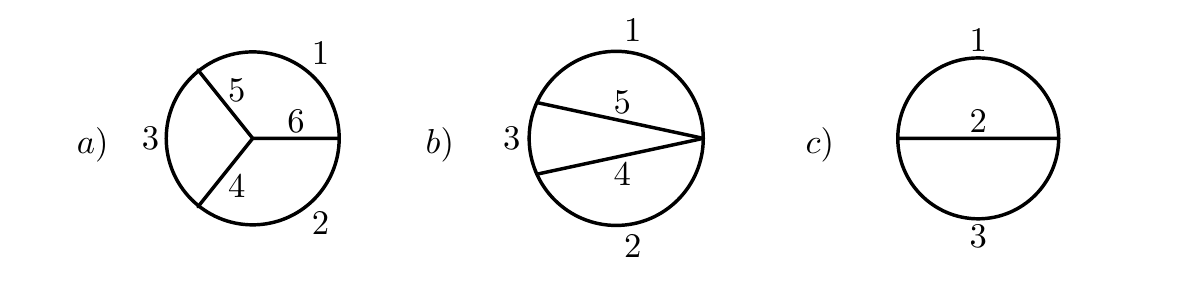}
\caption{\label{fig:topos} Topologies of three vacuum graphs appearing in the determination of the 3-loop EoS of cold quark matter \cite{Kurkela:2009gj} and discussed in the main text.}
\end{center}
\end{figure}

To illustrate this point as well as our notation for the sets $S_r\in S$ introduced in section \ref{sec:proof}, let us first consider the topology $a$ of fig.~\ref{fig:topos}. According to our notation, we then have $P_a=\{1,2,3,4,5,6\}$ as well as
\begin{eqnarray}
S_a&=&\{\{1,2,3\},\{1,2,4\},\{1,2,5\},\{1,3,4\},...\},
\end{eqnarray}
where the only sets of three indices missing from $S$ are $\{1,2,6\},\, \{2,3,4\},\,\{1,3,5\}$, and $\{4,5,6\}$, corresponding to the four three-vertices of the graph (cf.~fig.~\ref{fig:topos}). Writing down the first few terms of eq.~(\ref{Iresmu0}), we similarly obtain:
\begin{eqnarray}
I_a&=&\frac{1}{(2E_1)(2E_2)(2E_3)}\frac{1}{(iE_1-iE_2)^2+E_6^2}\frac{1}{(iE_2-iE_3)^2+E_4^2}\frac{1}{(iE_3-iE_1)^2+E_5^2}\nonumber \\
&+&\frac{1}{(2E_1)(2E_2)(2E_4)}\frac{1}{(iE_1-iE_2)^2+E_6^2}\frac{1}{(iE_2-iE_4)^2+E_3^2}\frac{1}{(iE_2-iE_1-iE_4)^2+E_5^2}\nonumber \\
&+&\cdots\, .
\end{eqnarray}
For the diagrams $b$ and $c$, the corresponding supersets $S$ read 
\begin{eqnarray}
S_b&=&\{\{1,2,3\},\{1,2,4\},\{1,2,5\},\{1,3,4\},\{1,4,5\},\{2,3,5\},\{2,4,5\},\{3,4,5\}\}\,, \nonumber \\
S_c&=&\{\{1,2\},\{1,3\},\{2,3\}\}\,.
\end{eqnarray}

\section{On the uniqueness of the cutting rules \label{sec:ind}}

In this appendix, we provide a detailed argument for the final form of the cutting rules, eq.~(\ref{ASr2}), starting from the earlier result of eq.~(\ref{ASr}). To this end, we define the function 
\begin{eqnarray}
J(S_r) &\equiv& 2^M \prod_{i=1}^M E_i \times I(S_r) \nonumber \\
&=& 2^M \prod_{i=1}^M E_i \times\sum_{S_{{r'}}\in S}\;\prod_{i\in S_{{r'}}} \frac{1}{2E_i}\, \tilde{A}_{S_r}(S_{{r'}})\Big{|}_{\{q_0^i=iE_i\}} \nonumber \\
&\equiv& \sum_{S_{r'} \in \mathcal S}\prod_{\alpha \in P \backslash S_{r'}} \left. \frac{E_{\alpha}(S_{r'})\theta_{r{r'}}}{(r_0^{\alpha}(S_{r'}))^2+E_{\alpha}^2(S_{r'})} \right|_{\{q_0^i=iE_i,\,i\in S_{r'}\}}\,.
\end{eqnarray}
where we denote by $\theta_{rr'}$ dimensionless coefficients composed of $\theta$-functions that may in principle depend both on $S_r$ and $S_{r'}$. From eq.~(\ref{ASr}) we know that $\theta_{rr}=1$ for all $r$, and we shall now show that the independence of the $E_\alpha$ implies that $\theta_{rr'}=1$ even when $r\neq r'$.

To achieve the above goal, we choose another $S_{r''}\in S$ and multiply the function $J(S_r)$ by the product $\prod_{\beta \in P \backslash S_{r''}}  E_\beta(S_{r''})$, after which we take the limit where the $E_\beta$ approach infinity:
\begin{eqnarray}
&&\lim_{E_{\beta}\to\infty} \prod_{\beta \in P \backslash S_{r''}}   E_\beta(S_{r''})\times J(S_r) \nonumber \\
&=& \sum_{S_{r'} \in \mathcal S} \lim_{E_{\beta} \to \infty} \prod_{\alpha \in P \backslash S_{r'}}
\prod_{\beta \in P \backslash S_{r''}} 
\left. \frac{E_{\alpha}(S_{r'})E_{\beta}(S_{r''})\theta_{rr'}}{(r_0^{\alpha}(S_{r'}))^2+E_{\alpha}^2(S_{r'})} \right|_{\{q_0^i=iE_i,\,i\in S_{r'}\}}\,.
\end{eqnarray}
At this point, we notice that in those terms of the sum where $r'\neq r''$ we have at least one index $\beta$ that belongs to the set $S_{r'}$. The corresponding $E_\beta$ thus appears only linearly in the numerator, but quadratically in the denominator. This implies that the corresponding limit must tend to 0, leaving us with 
\begin{eqnarray}
&&\lim_{E_{\beta}\to\infty} \prod_{\beta \in P \backslash S_{r''}}   E_\beta(S_{r''})\times J(S_r) \nonumber \\
&=& \lim_{E_{\beta} \to \infty} 
\prod_{\beta \in P \backslash S_{r''}} 
\left. \frac{E_{\beta}^2(S_{r''})\theta_{rr''}}{(r_0^{\beta}(S_{r''}))^2+E_{\beta}^2(S_{r''})} \right|_{\{q_0^i=iE_i,\,i\in S_{r''}\}}\; = \; \theta_{rr''}\,.
\end{eqnarray}
Knowing that $J(S_r)$ must be independent of $r$ --- just as $I(S_r)$ is --- we conclude from here that $\theta_{rr'} = \theta_{r'r'}=1$ and hence  $\tilde{A}_{S_r}(S_{r'})=\tilde{A}_{S_{r'}}(S_{r'})$, which is what we wanted to show.

\section{Zero-temperature limit of a fermionic Matsubara contour \label{sec:matsu}}

In this appendix, we demonstrate that the use of an infinitesimal temperature as a regulator of finite-$\mu$ Feynman graphs leads to the handling of divergences along the $p_0$ integration contour in terms of a principal value prescription. 

To begin, we consider a generic fermionic Matsubara sum, denoted by $T \sum_n h\left(i \omega_n\right)$, where $\omega_n=(2n+1)\pi T$ and the chemical potential resides in the function $h(z)$ that is taken to vanish sufficiently rapidly at large $|z|$. As usual, we assume that this function may be analytically continued to a meromorphic function $h:\mathbb{C}\rightarrow\mathbb{C}$. Letting $\varepsilon>0$, we then denote by $\Omega$ the $\varepsilon$-strip $\Omega\approx\left(-\varepsilon,\varepsilon\right)\times\mathbb{R}$, noting that if $h$ is holomorphic on $\Omega$, we may evaluate the sum by multiplying $h$ by an appropriately normalized Fermi distribution function that has poles at $\omega=i\omega_n$. This leads to the usual integral representation
\begin{eqnarray}
T \sum_{\left\{ \omega_n\right\} }h\left(i \omega_n\right)&=&-\sum_{\left\{ \omega_n\right\} }\mathrm{Res}\left[h\left(z\right)n_F\left(z\right)|z=i\omega_n\right] = \frac{1}{2\pi i}\lim_{\varepsilon\rightarrow0^{+}}\ointop_{\Gamma_{\varepsilon}}dz\,h\left(z\right)n_F\left(z\right), \label{matsu1}
\end{eqnarray}
where $n_F(z)\equiv 1/(e^{z/T}+1)$ and $\Gamma_{\varepsilon}$ denotes a \textit{clockwise} rectangular contour whose long sides lie on $\left\{ -\varepsilon\right\} \times\mathbb{R}$ and $\left\{ \varepsilon\right\} \times\mathbb{R}$, respectively. As the horizontal sides of the rectangle produce vanishing contributions, we may equivalently close the vertical contours by two infinite semicircles on the left and right halves of the complex plane. 

Proceeding to the zero-temperature limit, we may easily take $\varepsilon\rightarrow0^{+}$, which makes the two vertical contours pinch together. Taking advantage of the relation $\lim_{T\rightarrow0^{+}}n_F(z)=\theta(-{\rm Re}\,z)$ then leads to the simple result $\intop_{-i\infty}^{i\infty}\frac{dz}{2\pi i}\,h\left(z\right)$, where it is customary to redefine the integration variable as $z=i\tilde{z}$ so that we obtain a Euclidean signature integral along the real axis. This is a well-known result that we used as the starting point in our derivation of the cutting rules. Unfortunately however, not all physically interesting functions $h$ are holomorphic on the strip $\Omega$, as they may develop poles along the imaginary axis. This means that special care must be applied when proceeding to the $T\rightarrow0^{+}$ limit in the Matsubara sum, as we shall presently demonstrate.

\begin{figure}[t]
\begin{center}
\includegraphics[width=1\linewidth]{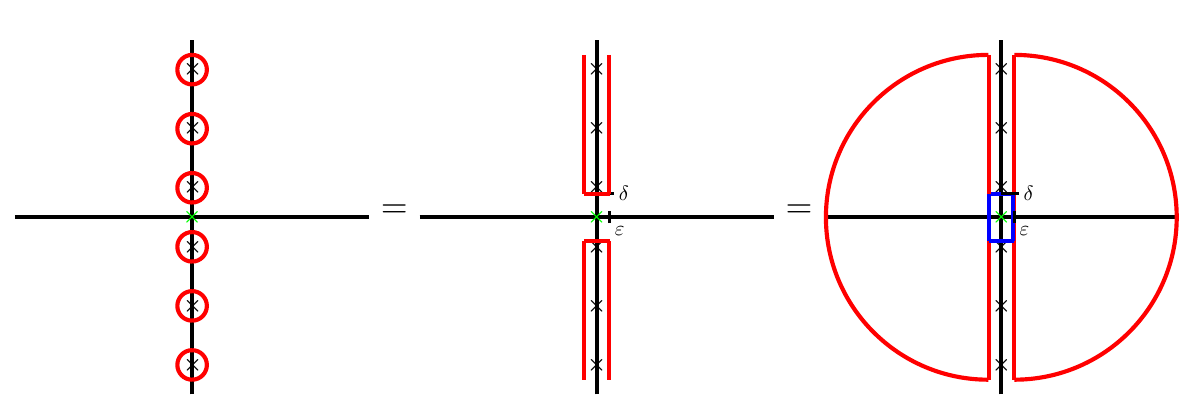}
\caption{\label{fig:matsu} An illustration of the procedure by which we convert a Matsubara sum into a contour integral when the summand has a pole at $z=0$. The red color represents the original integration contour $\Gamma_\epsilon$, while the blue rectangle stands for $\partial\bar{\Psi}_{\varepsilon,\delta}$. }
\end{center}
\end{figure}

Let us now choose $\delta\in\left(0,\frac{\pi T}{2}\right)$, and make the simplifying assumption that the only problematic pole of the function $h$ resides at the origin, $z=0$.\footnote{Other isolated poles not coinciding with the imaginary Matsubara frequencies can be easily removed in the same way, so this is not a restriction.} In this case, the integral over $\Gamma_{\varepsilon}$ has an unphysical contribution not present in the original Matsubara sum that can be removed by integrating the function $h(z) n_F(z)$ clockwise over the boundary of the rectangle $\Psi_{\varepsilon,\delta}=\left(-\varepsilon,\varepsilon\right)\times\left(-i\delta,i\delta\right)$ (note that this function is holomorphic on $\bar{\Psi}_{\varepsilon,\delta}\backslash\left\{ 0\right\}$,
since $\underset{z\in\bar{\Psi}_{\varepsilon,\delta}}{\sup}\left|\mathrm{Im}z\right| \leq \frac{\pi T}{2}<\pi T$). This yields as the generalization of eq.~(\ref{matsu1})
\begin{eqnarray}
T \sum_{\left\{ \omega_n\right\} }h\left(i\omega_n\right)&=&\frac{1}{2\pi i}\lim_{\delta\rightarrow0^{+}}\lim_{\varepsilon\rightarrow0^{+}}\bigg\{ \ointop_{\Gamma_{\varepsilon}}dz\,h\left(z\right)n_F\left(z\right)-\ointop_{\partial\bar{\Psi}_{\varepsilon,\delta}}dz\,h\left(z\right)n_F\left(z\right)\bigg\}\,, \label{matsu2}
\end{eqnarray}
which we depict in fig.~\ref{fig:matsu} and where we have at the end taken the limit that $\delta$, too, tends to zero. 

Let us now inspect the form of eq.~(\ref{matsu2}) in detail. Considering first the horizontal sides of the rectangular integral, we note that the integrand is regular along them, so that they independently vanish in the $\epsilon\to 0^+$ limit,
\begin{eqnarray}
\lim_{\varepsilon\rightarrow0^{+}}\intop_{\left(-\varepsilon,\pm i\delta\right)}^{\left(\varepsilon,\pm i\delta\right)}dz\,h\left(z\right)n_F\left(z\right)=0\;\;\;\;\forall\ T>0,\ \delta\in\Big(0,\frac{\pi T}{2}\Big)\,.
\end{eqnarray}
At the same time, the arc integrals in $\Gamma_{\varepsilon}$ are unchanged (i.e.~they still vanish at infinity), so for the first term in eq.~(\ref{matsu2}) we are left with the usual result $\intop_{-i\infty-\varepsilon}^{i\infty-\varepsilon}dz\,h\left(z\right)n_F\left(z\right)-\intop_{-i\infty+\varepsilon}^{i\infty+\varepsilon}dz\,h\left(z\right)n_F\left(z\right)$. Subtracting from here the vertical sides of the second term of eq.~(\ref{matsu2}) yields then $\intop_{-i\infty\pm\varepsilon}^{-i\delta\pm\varepsilon}dz\,h\left(z\right)n_F\left(z\right)+\intop_{i\delta\pm\varepsilon}^{i\infty\pm\varepsilon}dz\,h\left(z\right)n_F\left(z\right)$, so that in total, we obtain
\begin{eqnarray}
T\sum_{\left\{ \omega_n\right\} }h\left(i\omega_n\right)
=\frac{1}{2\pi i} \lim_{\delta\rightarrow0^{+}}\lim_{\varepsilon\rightarrow0^{+}}&\,& \left[ \intop_{-i\infty-\varepsilon}^{-i\delta-\varepsilon}dz\,h\left(z\right)n_F\left(z\right)+\intop_{i\delta-\varepsilon}^{i\infty-\varepsilon}dz\,h\left(z\right)n_F\left(z\right) \right. \nonumber\\
&-& \left. \intop_{-i\infty+\varepsilon}^{-i\delta+\varepsilon}dz\,h\left(z\right)n_F\left(z\right)-\intop_{i\delta+\varepsilon}^{i\infty+\varepsilon}dz\,h\left(z\right)n_F\left(z\right) \right].\;\;\;\;
\end{eqnarray}
Taking now advantage of the fact that the integrand is regular along the integration contour, we may proceed to the $T\rightarrow0^{+}$ limit in the usual manner. This gives as the zero-temperature limit of the Matsubara sum
\begin{eqnarray}
T\sum_{\left\{ \omega_n\right\} }h\left(i\omega_n\right)&\overset{T\rightarrow0^{+}}{\rightarrow}&\frac{1}{2\pi i}\lim_{\delta\rightarrow0^{+}}\bigg\{ \intop_{-i\infty}^{-i\delta}dz\,h\left(z\right)+\intop_{i\delta}^{i\infty}dz\,h\left(z\right)\bigg\} \equiv\mathcal{P}\int_{-\infty}^{\infty}\frac{d\tilde{z}}{2\pi}\, h\left(i\tilde{z}\right)\,,\;\;\;\; \label{prinval}
\end{eqnarray}
where we have arrived at a principal value type integral.  This result implies that the correct starting point for the derivation of the cutting rules is to define the integration measure as in eq.~(\ref{principal}).


\begin{thebibliography}{99}


  
\bibitem{deForcrand:2010ys}
  P.~de Forcrand,
  PoS LAT {\bf 2009} (2009) 010
  [arXiv:1005.0539 [hep-lat]].
  
\bibitem{Machleidt:2011zz}
  R.~Machleidt and D.~R.~Entem,
  Phys.\ Rept.\  {\bf 503} (2011) 1
  doi:10.1016/j.physrep.2011.02.001
  [arXiv:1105.2919 [nucl-th]].
  
\bibitem{Kraemmer:2003gd}
  U.~Kraemmer and A.~Rebhan,
  Rept.\ Prog.\ Phys.\  {\bf 67} (2004) 351
  doi:10.1088/0034-4885/67/3/R05
  [hep-ph/0310337].
  
\bibitem{Hebeler:2013nza}
  K.~Hebeler, J.~M.~Lattimer, C.~J.~Pethick and A.~Schwenk,
  Astrophys.\ J.\  {\bf 773} (2013) 11
  doi:10.1088/0004-637X/773/1/11
  [arXiv:1303.4662 [astro-ph.SR]].
  
  
\bibitem{Kurkela:2014vha}
  A.~Kurkela, E.~S.~Fraga, J.~Schaffner-Bielich and A.~Vuorinen,
  Astrophys.\ J.\  {\bf 789} (2014) 127
  doi:10.1088/0004-637X/789/2/127
  [arXiv:1402.6618 [astro-ph.HE]].
  
\bibitem{Fraga:2015xha}
  E.~S.~Fraga, A.~Kurkela and A.~Vuorinen,
  Eur.\ Phys.\ J.\ A {\bf 52} (2016) no.3,  49
  doi:10.1140/epja/i2016-16049-6
  [arXiv:1508.05019 [nucl-th]].
  
\bibitem{Freedman:1976ub}
  B.~A.~Freedman and L.~D.~McLerran,
  Phys.\ Rev.\ D {\bf 16} (1977) 1169.
  doi:10.1103/PhysRevD.16.1169
  
\bibitem{Vuorinen:2003fs}
  A.~Vuorinen,
  Phys.\ Rev.\ D {\bf 68} (2003) 054017
  doi:10.1103/PhysRevD.68.054017
  [hep-ph/0305183].
  
  
\bibitem{Kurkela:2009gj}
  A.~Kurkela, P.~Romatschke and A.~Vuorinen,
  Phys.\ Rev.\ D {\bf 81} (2010) 105021
  doi:10.1103/PhysRevD.81.105021
  [arXiv:0912.1856 [hep-ph]].
  
\bibitem{Fraga:2004gz}
  E.~S.~Fraga and P.~Romatschke,
  Phys.\ Rev.\ D {\bf 71} (2005) 105014
  doi:10.1103/PhysRevD.71.105014
  [hep-ph/0412298].
  
\bibitem{Kurkela:2016was}
  A.~Kurkela and A.~Vuorinen,
  Phys.\ Rev.\ Lett.\  {\bf 117} (2016) no.4,  042501
  doi:10.1103/PhysRevLett.117.042501
  [arXiv:1603.00750 [hep-ph]].
  
\bibitem{Kajantie:2002wa}
  K.~Kajantie, M.~Laine, K.~Rummukainen and Y.~Schroder,
  Phys.\ Rev.\ D {\bf 67} (2003) 105008
  doi:10.1103/PhysRevD.67.105008
  [hep-ph/0211321].
  
\bibitem{DiRenzo:2006nh}
  F.~Di Renzo, M.~Laine, V.~Miccio, Y.~Schroder and C.~Torrero,
  JHEP {\bf 0607} (2006) 026
  doi:10.1088/1126-6708/2006/07/026
  [hep-ph/0605042].
  
  \bibitem{nextpaper}
  I.~Ghisoiu, T.~Gorda, A.~Kurkela, P.~Romatschke, and A.~Vuorinen, In preparation.
  

\bibitem{Gynther:2007bw}
  A.~Gynther, M.~Laine, Y.~Schroder, C.~Torrero and A.~Vuorinen,
  JHEP {\bf 0704} (2007) 094
  doi:10.1088/1126-6708/2007/04/094
  [hep-ph/0703307 [HEP-PH]].
  
\bibitem{Gynther:2009qf}
  A.~Gynther, A.~Kurkela and A.~Vuorinen,
  Phys.\ Rev.\ D {\bf 80} (2009) 096002
  doi:10.1103/PhysRevD.80.096002
  [arXiv:0909.3521 [hep-ph]].

\bibitem{Andersen:2000zn}
  J.~O.~Andersen, E.~Braaten and M.~Strickland,
  Phys.\ Rev.\ D {\bf 62} (2000) 045004
  doi:10.1103/PhysRevD.62.045004
  [hep-ph/0002048].
 
  
\bibitem{Dashen:1969ep}
  R.~Dashen, S.~K.~Ma and H.~J.~Bernstein,
  Phys.\ Rev.\  {\bf 187} (1969) 345.
  doi:10.1103/PhysRev.187.345

 
\bibitem{Bugrii:1995vn}
  A.~I.~Bugrii and V.~N.~Shadura,
  hep-th/9510232.
  
\bibitem{Frenkel:1992az}
  J.~Frenkel, A.~V.~Saa and J.~C.~Taylor,
  Phys.\ Rev.\ D {\bf 46} (1992) 3670.
  doi:10.1103/PhysRevD.46.3670
   
\bibitem{Laine:2006cp}
  M.~Laine and Y.~Schroder,
  Phys.\ Rev.\ D {\bf 73} (2006) 085009
  doi:10.1103/PhysRevD.73.085009
  [hep-ph/0603048].
  
  
  
\bibitem{Kapusta:2006pm}
  J.~I.~Kapusta and C.~Gale,
  ``Finite-temperature field theory: Principles and applications.''

  
\end{thebibliography}
\end{document}